\documentclass[12pt,onecolumn,draftcls,journal,letterpaper]{IEEEtran}

\usepackage{amsthm}
\usepackage[usenames,dvipsnames,svgnames,table]{xcolor}
\usepackage{amssymb,amsmath,color} 
\usepackage{graphicx} 
\usepackage{epsfig}
\usepackage{algorithm}
\usepackage[noend]{algpseudocode}

\usepackage{pgfplots}
\pgfplotsset{compat=newest}

\makeatletter
\newcommand{\StatexIndent}[1][3]{%
  \setlength\@tempdima{\algorithmicindent}%
  \Statex\hskip\dimexpr#1\@tempdima\relax}
\makeatother

\usepackage{psfrag}

\newcommand{\real}{{\mathbb{R}}}

\newcommand{\diag}{\operatorname{diag}} 

\newcommand{\CC}{\mathcal{C}} 
 \newcommand{\FF}{\mathcal{F}}
\newcommand{\GG}{\mathcal{G}} 
 
 \newcommand{\PP}{\mathcal{P}}
 
\newcommand{\TT}{\mathcal{T}}

\renewcommand{\d}{\mathrm{d}}

 \newcommand{\subj}{\text{subj. to}}
\newcommand{\subject}{\text{subject to}} 
\newcommand{\minimize}{\text{minimize}}

\newcommand{\argmin}{\mathop{\rm argmin}}

\newtheorem{theorem}{Theorem}[section]

\newtheorem{remark}[theorem]{Remark}

\newtheorem{assumption}[theorem]{Assumption}
 
\newcommand\oprocendsymbol{\hbox{$\square$}}
\newcommand\oprocend{\relax\ifmmode\else\unskip\hfill\fi\oprocendsymbol}

\usepackage{tikz}
\usetikzlibrary{shapes,arrows}

\newcommand{\FB}[1]{{\color{cyan} #1}}

\def \unf/{u}
\def \fcd/{f}

\begin{document}

\title{Final-State Constrained Optimal Control via a Projection Operator
  Approach}






\author{Ivano Notarnicola$^{1}$, Florian A. Bayer$^{2}$, Giuseppe Notarstefano$^{1}$, and Frank Allg\"ower$^{2}$
\thanks{$^{1}$Ivano Notarnicola and Giuseppe Notarstefano are with the Department of Engineering, Universit\`a del Salento, Lecce, Italy,
\tt\small name.lastname@unisalento.it}
\thanks{$^{2}$Florian Bayer and Frank Allg\"ower are
  with the Institute for Systems Theory and Automatic Control,
  University of Stuttgart, 70550 Stuttgart, Germany,
  {\tt\small \{bayer, allgower\}@ist.uni-stuttgart.de}}%
\thanks{
  This result is part of a project that has received funding from the European Research 
  Council (ERC) under the European Union's Horizon 2020 research and innovation 
  programme (grant agreement No 638992 - OPT4SMART).
  \newline \indent  
  F. Bayer and F. Allg\"ower would like to thank the German Research Foundation (DFG) for financial support within the 
  Cluster of Excellence in Simulation Technology (EXC 310/2) at the University of Stuttgart.
}}

\maketitle 


\begin{abstract}
  In this paper we develop a numerical method to solve nonlinear optimal control
  problems with final-state constraints. Specifically, we extend the PRojection
  Operator based Netwon's method for Trajectory Optimization (PRONTO), which was
  proposed by Hauser for unconstrained optimal control problems. While in the
  standard method final-state constraints can be only approximately handled by
  means of a terminal penalty, in this work we propose a methodology to meet the
  constraints exactly. Moreover, our method guarantees recursive feasibility of
  the final-state constraint. This is an appealing property especially in
  realtime applications in which one would like to be able to stop the
  computation even if the desired tolerance has not been reached, but still
  satisfy the constraints.  Following the same conceptual idea of PRONTO, the
  proposed strategy is based on two main steps which (differently from the
  standard scheme) preserve the feasibility of the final-state constraints: (i)
  solve a quadratic approximation of the nonlinear problem to find a descent
  direction, and (ii) get a (feasible) trajectory by means of a feedback law
  (which turns out to be a nonlinear projection operator).
  To find the (feasible) descent direction we take advantage of final-state
  constrained Linear Quadratic optimal control methods, while the second step is
  performed by suitably designing a constrained version of the trajectory
  tracking projection operator.
  The effectiveness of the proposed strategy is tested on the optimal state
  transfer of an inverted pendulum.





\end{abstract}


\section{Introduction}
\label{sec:introduction}
Optimal control problems (OCPs) are an active field of research in the controls
community since they may arise in many application areas as, e.g., Process
Control, Robotics, Aerospace and Automotive.
Throughout the last decades, many different approaches have been presented to
solve these problems. A possible classification of these methods has been given
in \cite{Diehl06Book}: (i) Dynamic programming, (ii) Indirect Methods, and (iii)
Direct methods. While methods in the first class solve the OCP by finding
optimal input segments using the Principle of Optimality (see, e.g.,
\cite{Bertsekas05}, or \cite{Bryson75}), the ones in the second area are based
on solving the necessary conditions for optimality using a (two-point) boundary
value problem, which can be solved by means of calculus of variations
(\cite{Kirk70}, \cite{Sage68}) or Pontryagin's Maximum Principle
(\cite{Pontryagin62}, \cite{Liberzon12}).
The third direction is the most
investigated and simplifies the OCP by parameterizing the control.  According to
the way the dynamics is handled, these methods are classified into fully
discretized (or collocation) methods (see, e.g.,~\cite{Cervantes98}) and direct
shooting methods, where the dynamics are included by some integration scheme
(see, e.g., ~\cite{Bock84}).  A detailed overview over Direct methods can, for
example, be found in~\cite{Betts01}.

Of special interest for our paper is the PRojection Operator based Newton method
for Trajectory Optimization (PRONTO) which was introduced in \cite{Hauser02},
see also \cite{Hauser98}.
In contrast to many other approaches solving optimal
control problems, this method is able to guarantee feasibility of the dynamics
after each iteration of the
underlying Newton method using a
``projection operator'' defined by a feedback, closed-loop system. 
According to the classification in \cite{Diehl06Book} this can be seen as
a combination of shooting and collocation.

This method was designed to handle unconstrained optimal control problems (and
extended to input-constrained problems in \cite{Saccon08}), 
considering final-state
constraints only approximately by means of a final
penalty.
Matching exactly final-state constraints is of interest in many
control applications.  This is the case, for example, in the field of hybrid
systems, that is, systems that consist of continuous and discrete event dynamics
(see, e.g.,~\cite{Goebel12} and the references therein). Discontinuous jumps of
continuous states may occur when the system state traverses a certain region of
the state space. This demands for an exact satisfaction of constraints on the
final state.
Another field where this is of interest is the field of Model Predictive
Control (MPC) (see, e.g., \cite{Rawlings09} and the references therein). In
MPC, the system is controlled by means of repeatedly solving a finite-horizon
OCP.
In many approaches within MPC, convergence and stability can be guaranteed if a certain terminal condition is satisfied. 
This leads to the need of an algorithm being able to handle final state constraints.

A first approach to solve the nonlinear transfer problem was introduced
in~\cite{hauser2003computation}. In there, the terminal constraint was
satisfied asymptotically by iteratively choosing a terminal reference until the
actual final state matches the target one.

The contribution of this paper is twofold. First, we introduce a new projection
operator, inspired by the one presented in~\cite{Hauser98}, such that not only
the dynamics, but also the terminal constraint is satisfied after each iteration
of the optimization algorithm. We reformulate the constrained projection as a
root-finding of an infinite dimensional functional, which can be accomplished by
means of a Newton root-finding in Banach spaces. Then, based on this new
projection operator, as main contribution we propose an optimal control method
solving final-state constrained problems which shows recursive feasibility. The
proposed algorithm consists of two steps. First, a feasible descent direction
is determined using a quadratic approximation of the nonlinear problem. The
descent direction is chosen such that the mismatch on the final state is zero.
Second, the perturbed curve is projected on the feasible manifold such that the
\emph{dynamics and the terminal constraint} are satisfied.

An interesting feature of the proposed algorithm is that it is amenable to
realtime, fast MPC schemes. Indeed, in many applications one may not be able to
run the algorithm until convergence is achieved with a desired tolerance. Due to
a reduced computation time it could be that a (much) shorter number of
iterations can be run. Since feasibility of both the dynamics and the final
state-constraint are guaranteed at each iteration one can stop the computation
and still get a feasible trajectory.
 
The paper is organized as follows. In Section~\ref{sec:problem_setup} we
introduce the problem setup and recall how to solve final-state constrained
linear quadratic optimal control problems. PRONTO is introduced in
Section~\ref{subsec:pronto}. Our new final-state constrained PRONTO is
presented in Section~\ref{sec:fs_pronto} and a numerical simulation for the
optimal state-transfer of an inverted pendulum is given in
Section~\ref{sec:simulations}.

\paragraph*{Notation}
Given a smooth vector field $f(x,u)$, we denote by $f_x (\bar{x}, \bar{u})$ its
derivative with respect to $x$ evaluated at $(\bar{x}, \bar{u})$, and,
consistently, by $f_u$ its derivative with respect to $u$. For the curve
$\xi = (x(\cdot),u(\cdot))$, we introduce the projections $\pi_{1} = [I~0]$ and
$\pi_{2} = [0~I]$ such that $x(\cdot) = \pi_1 \xi$ and $u(\cdot) = \pi_2 \xi$.
Given a functional $\GG : X \to \real$, with $X$ a Banach space, and a point
$\xi \in X$, we denote by $D \GG(\xi)$ the first Fr\'echet derivative of $\GG$
evaluated at $\xi$, and, consistently, by $D^2 \GG(\xi)$ its second Fr\'echet
derivative, \cite{Zeidler95}.

\section{Problem Setup and Preliminaries}
\label{sec:problem_setup}
In this paper we consider a final-state constrained optimal control
problem. That is, we aim at finding a trajectory of a dynamical system that
minimizes a given objective functional while satisfying an initial and a
terminal constraint.
%
%
Formally, we consider the problem
\begin{align}
\begin{split}
  \underset{(x(\cdot),u(\cdot))}{\text{minimize}} \: & \: \int_0^T \ell(x(\tau), u(\tau)) \,\d \tau
  \\
  \subject \:  & \: \dot{x}(t) = f(x(t),u(t))
  \\
  & \: x(0) = x_0, ~x(T) = x_T,
\end{split}
\label{eq:nonlin_state_transf}
\end{align}
where $\ell : \real^n \times \real^m \to \real$ is the running cost,
$f : \real^n \times \real^m \to \real^n$ is the nonlinear vector field
describing the control system, and $x_0 \in \real^n$ and $x_T \in \real^n$ are
the initial and final fixed states respectively. We assume $\ell$ and $f$ to be
$\CC^2$ functions. 
Notice that in the rest of the paper, for the sake of brevity, we will omit the
dimensions of the quantities when it will be clear from the equations.



Before stating the main assumptions for problem \eqref{eq:nonlin_state_transf},
we recall some notation that will be also useful in the rest of the paper.
Consider the Hamiltonian of \eqref{eq:nonlin_state_transf} given by
\begin{align}
  \! H\! (x(t),p(t),u(t)) \! := \! \ell (x(t), u(t)) \! +\! p(t)^T \! f(x(t),u(t)),
  \label{eq:hamiltonian}
\end{align}
where $p(\cdot)$ is the costate. Then, for $\xi = (\bar x(\cdot),\bar u(\cdot))$ define
\begin{align}
  \! q(\xi) \!\cdot\! (\zeta, \zeta) \!:=\! \int_0^T\!
  \begin{bmatrix}
    z(\tau) \\ v(\tau)
  \end{bmatrix}^T
  \!\!
  \begin{bmatrix}
    H_{xx}(\tau) &\!\!\! H_{xu}(\tau)\\
    H_{ux}(\tau) &\!\!\! H_{uu}(\tau)
  \end{bmatrix}
  \!\!
  \begin{bmatrix}
    z(\tau) \\ v(\tau)
  \end{bmatrix} \!
  \d\tau,
  \label{eq:q}
\end{align}
where $\zeta = (z(\cdot),v(\cdot))$ is a (state-input) curve representing a
variation from $\xi$, while $H_{xx}(t)$, $H_{xu}(t)$ and $H_{uu}(t)$ denote the
appropriate second derivative of the $H$ evaluated along the extremal
state-control-costate trajectory, e.g.,
$H_{xx}(t) = H_{xx}(\bar{x}(t), \bar{p}(t), \bar{u}(t))$.

Given a dynamical system $\dot x = f(x,u)$, $x(0) = x_0$, we say that a state-input curve
$\xi = (\bar x(t),\bar u(t))$ is a trajectory of the system if it satisfies the
dynamics, i.e., $\dot{\bar x}(t) = f( \bar x(t), \bar u(t))$ for all $t\in[0,T]$ and
$\bar x(0) = x_0$. We denote the (infinite-dimensional) manifold of all system
trajectories by $\TT$, so that we write $\xi \in \TT$.

Given a trajectory $\xi = (\bar x(t),\bar u(t))$, we denote by $T_\xi \TT$ the
manifold of curves $\zeta = (z(\cdot),v(\cdot))$ satisfying the linearized
dynamics
\begin{align}
  \dot z = f_x (\bar x(t),\bar u(t)) z +f_u(\bar x(t),\bar u(t))v
\label{eq:linearization}
\end{align}
with $z(0) = 0$ and for $v(\cdot) \in L_2$. We say that $T_\xi \TT$ is the
tangent space of the trajectory manifold at $\xi$.

\begin{assumption}[Linear controllability]
  The system $\dot x = f(x,u)$ is linearly controllable around any
  trajectory. That is, for any $(\bar{x}(\cdot), \bar{u}(\cdot))$ defined on
  $[0,T]$, the linearized system~\eqref{eq:linearization}
  is controllable over $[0,T]$. 
\label{ass:linear_ctrl}
\end{assumption}

\begin{assumption}[Second Order Sufficiency]
  Given a trajectory $\xi\in\TT$, the Hamiltonian $H$ satisfies
  $H_{uu}(t) \ge r_0I$ for $t \in [0,T]$ and some $r_0 > 0$, and the quadratic
  functional $q$ is positive-definite\footnote{See, e.g., \cite{Zeidler95} for
  the definition of positive definite functional.} on $T_\xi\TT$. \oprocend
\label{ass:SSC}
\end{assumption}

\begin{theorem}[{\cite[Theorem~$2.1$]{hauser2003computation}}]
  Let $\xi = (x(\cdot), u(\cdot))$ be a stationary trajectory
  of~\eqref{eq:nonlin_state_transf} with corresponding costate trajectory
  $p(\cdot)$. Suppose that Assumption~\ref{ass:SSC} hold at $\xi$. If the system
  is linearly controllable around $\xi$, then $\xi$ is an isolated local minimum
  of~\eqref{eq:nonlin_state_transf}. \oprocend
\end{theorem}

\begin{remark}
  Assumption~\ref{ass:linear_ctrl} not only is a sufficient condition for the
  theorem above, but also guarantees that the algorithm we propose will be
  solvable at each iteration.
  \oprocend 
\end{remark}

\subsection{Linear Quadratic (LQ) optimal state transfer problem}
\label{subsec:fscLQR}
We start by considering a special version of
problem~\eqref{eq:nonlin_state_transf} in which the cost is quadratic and the
dynamics is linear and time-varying, i.e., we consider the problem
\begin{align}
  \label{eq:lin_state_transf}
  \begin{split}
  \underset{(x(\cdot),u(\cdot))}{\text{minimize}} \: & \: \int_0^T
  a(\tau)^T x(\tau) + b(\tau)^T u(\tau)
  \\ 
  & \: +
  \dfrac{1}{2}
  \begin{bmatrix}
    x(\tau) \\ u(\tau)
  \end{bmatrix}^T
  \begin{bmatrix}
    Q(\tau) & S(\tau)\\
    S(\tau)^T & R(\tau)
  \end{bmatrix}
  \begin{bmatrix}
    x(\tau) \\ u(\tau)
  \end{bmatrix} 
  \d\tau
  \\[1ex]
  \subject \: & \: 
  \dot{x} = A(t)x + B(t)u, \: x(0)=x_0, \: x(T)=x_T,
  \end{split}
\end{align}
where we assume that $a(\cdot)$ and $b(\cdot)$ 
are piecewise continuous vectors, and $A(\cdot)$, 
$B(\cdot)$, $Q(\cdot) = Q(\cdot)^T$, $R(\cdot) = R(\cdot)^T$, 
and $S(\cdot)$ are piecewise continuous matrices with
$R(t) \geq r_0I$, $t \in [0,T]$, for some $r_0 > 0$.

\begin{remark}
  Problem~\eqref{eq:lin_state_transf} can be obtained as the linear-quadratic
  approximation of problem~\eqref{eq:nonlin_state_transf}. In particular, $A(\cdot)$
  and $B(\cdot)$ result from the linearization of the nonlinear dynamics $f$ at
  a given trajectory, while $Q$, $R$, $S$, $a$ and $b$ define the quadratic
  approximation of the nonlinear cost functional $\ell$ at the same trajectory. 
\oprocend
\end{remark}

\begin{theorem}[{\cite[Proposition~$1.1$]{hauser2003computation}}]
  If $(A(\cdot), B(\cdot))$ in \eqref{eq:lin_state_transf} describes a
  controllable linear time-varying system over $[0,T]$ and $q$ is positive
  definite on the space of the system trajectories, 
  then problem~\eqref{eq:lin_state_transf} has a unique solution. \oprocend
\end{theorem}

%

Next, we recall how to solve problem~\eqref{eq:lin_state_transf}. We start by
imposing the first-order necessary conditions of optimality. 

Setting to zero the first variation of the Hamiltonian with respect to $u$, we 
obtain the optimal feedback law 
\begin{align}
  \label{eq:u_feedback}
  u &= -R^{-1} [ \, S^T x + B^Tp + b \,].
\end{align}
By setting the first variations of the Hamiltonian with respect to $x$ and $p$ to zero 
and by using~\eqref{eq:u_feedback}, we obtain the following linear two-point boundary 
value problem
\begin{align}
  \label{eq:TPBVP}
  \hspace{-0.2cm}
  \begin{bmatrix}
    \dot{x} \\ \dot{p}
  \end{bmatrix}  
  \! \!= \!\!
  \begin{bmatrix}
    \tilde{A} & \hspace{-0.3cm} -BR^{-1}B^T\\
    -\tilde{Q} & \hspace{-0.3cm} -\tilde{A}^T
  \end{bmatrix}
  \!\!
  \begin{bmatrix}
    x \\ p
  \end{bmatrix}
  \! + \!
  \begin{bmatrix}
    -BR^{-1}b\\
    SR^{-1}b-a
  \end{bmatrix}\!,\!\!\!
  \begin{array}{l}
    x(0) \!=\! x_0\\
    p(T) \!=\! p_1
  \end{array}\!\!,
\end{align}
where $p(t)$ is the costate, $p_1$ is a boundary value to be determined,
$\tilde{A} := A-BR^{-1}S^T$ and $\tilde{Q} := Q - SR^{-1}S^T$.

%
%

It can be shown that $p$ and $x$ in \eqref{eq:TPBVP} are related via 
an affine relation, i.e.,
\begin{equation}
  p = P x + r.
  \label{eq:Ricc_transf}
\end{equation}
%
%
By defining the gain matrix
$K := R^{-1}(S^T+ B^TP)$,
%
the optimal input~\eqref{eq:u_feedback} results into the affine
feedback law
%
  $u = -Kx - R^{-1}(B^Tr+b).$
%
%
%
Then, equation \eqref{eq:TPBVP} can be decoupled by means of the sweep method,
\cite{Bryson75}, which leads to the following differential (Riccati) equations
%
\begin{align}
	\label{eq:P_dot}
  -\dot{P} &=A^TP + P A  -K^TRK+Q, && \quad P(T)=0\\
	\label{eq:r_dot}
  -\dot{r} &=  (A-BK)^T r - K^T b +a,  && \quad r(T)=p_1
\end{align}
where the boundary conditions follow from~\eqref{eq:Ricc_transf}. 

The above equations should be integrated to determine the optimal
control~\eqref{eq:u_feedback} and thus solve
problem~\eqref{eq:lin_state_transf}. However, the terminal vector $p_1$ is
still unknown.
Thus, we need to express explicitly the relation between $p_1$ and 
the terminal condition $x_T$. 
Plugging \eqref{eq:Ricc_transf} into the first equation of \eqref{eq:TPBVP}, we obtain 
\begin{align}
  \label{eq:x_closeloop}
  \dot{x} &= (A - BK) x - BR^{-1} ( B^Tr + b ), && x(0)=x_0.
\end{align}
Next, we observe that
\begin{align}
  \label{eq:terminal_x}
  x(T) = x_{\unf/}(T) + x_{\fcd/, b}(T) + x_{\fcd/, r}(T),
\end{align}
where $x_{\unf/}(T)$ is the unforced response of system \eqref{eq:x_closeloop} 
at time $t=T$, 
whereas $x_{\fcd/, b}(T)$ and $x_{\fcd/, r}(T)$ are the 
forced responses due to the inputs $BR^{-1} b$ and $BR^{-1} B^T r$, respectively. 

Focusing on $x_{\fcd/, r}(T)$, we note that it can be further split into two
contributions related, respectively, to the forced and unforced responses of
$r$. The latter contribution depends directly on $p_1$ and it can be shown that
equation~\eqref{eq:terminal_x} can be rewritten as
  $x(T) =   x_{\unf/}(T) + n(T) - W_c(T) p_1$, 
where $W_c(T)$ is the controllability Gramian matrix,
\begin{equation*}
  W_c(t) := \!\!\int_0^t \!\!\Phi_c (t,\tau) B(\tau)R(\tau)^{-1} B(\tau)^T \Phi_c(t,\tau)^T\,\d \tau,
\end{equation*}
evaluated at time $T$, with $\Phi_c$ being the state transition function
associated to closed-loop system with state matrix $A - BK$, while $n(T)$
denotes the terminal state of
\begin{align*}
  \dot{n} & = (A-BK) n - BR^{-1} ( B^T r_{\fcd/} + b ), && n(0)=0,
\end{align*}
where $r_{\fcd/}$ denotes the forced response of $r$, i.e., it solves
\eqref{eq:r_dot} with zero terminal condition.

To conclude, $p_1$ can be computed as
\begin{align*}
 p_1 =  W_c(T)^{-1} \left(x_T - x_{\unf/}(T) - n(T)\right).
\end{align*}

\section{Projection Operator Newton Method for Trajectory Optimization (PRONTO)}
\label{subsec:pronto}
PRONTO was introduced in~\cite{Hauser98} to solve the following finite-horizon
optimal control problem
\begin{align}
  \begin{split}
    \underset{(x(\cdot),u(\cdot))}{\minimize} \: & \: \int_0^T \ell (x(\tau),u(\tau)) \,\d\tau +m(x(T)) 
    \\
    \subject \: & \: \dot{x}(t) = f(x(t),u(t)), \hspace{0.5cm} x(0)=x_0,
  \end{split}
\label{eq:pronto_ocp}
\end{align}
which, differently from problem~\eqref{eq:nonlin_state_transf}, has a terminal
penalty $m : \real^n \to \real$ rather than a terminal constraint.

%


The key idea of PRONTO is to (i) convert the dynamically constrained
(infinite-dimensional) optimization problem into an unconstrained one by means
of a projection operator, and (ii) solve the unconstrained problem via an
infinite-dimensional Newton method.

We start recalling the projection operator, which is based on a trajectory tracking
feedback law. 

\subsection{The trajectory tracking nonlinear projection operator}
\label{subsec:proj_oper}
Suppose that $\xi := (\alpha(\cdot),\mu(\cdot))$ (defined on $t \geq 0$) is a
bounded state-input curve and let $\eta := (x(\cdot),u(\cdot))$ be the
trajectory determined by the nonlinear feedback system
\begin{align}
  \Bigg\{
  \begin{split}
  & \dot{x} (t) = f (x(t), u(t)),\hspace{1.4cm} x(0) = \alpha(0)
  \\
  & u(t) = \mu(t) + K(t)[\alpha(t) - x(t)].
  \end{split}
\label{eq:tracking_operator}
\end{align}

Under suitable conditions on $f$ and $K$, the feedback system in 
\eqref{eq:tracking_operator} defines a \emph{continuous} nonlinear projection operator 
$\PP : \xi = (\alpha (\cdot), \mu (\cdot) ) \mapsto \eta = (x(\cdot), u(\cdot))$.
%

The operator $\PP$ is a projection since $\PP = \PP\circ \PP $ on its
domain. Indeed, independent of $K$, if $\xi$ is a trajectory of $f$, then $\xi$
is a \emph{fixed point} of $\PP$, i.e., $\xi = \PP(\xi)$. As a consequence, 
a trajectory can be characterized in terms of the projection operator as
  $\xi \in \TT$ if and only if $\xi = \PP(\xi)$.
%
In \cite{Hauser98}, the authors have proven that the projection
operator $\PP$ is as smooth as $f$ and one can compute (and analyze) its
derivatives.
In particular, if $f$ is $\CC^1$, then the first derivative of the projection 
operator is the linear mapping $\zeta = (\beta(\cdot),\nu(\cdot)) \mapsto
D\PP(\xi)\cdot \zeta =  (z(\cdot),v(\cdot))$
%
defined by
\begin{align*}
\Bigg\{
\begin{split}
&\dot{z} (t) = f_x (x(t),u(t)) z(t) + f_u (x(t),u(t))  v(t),\: z(0) = 0\\
&v(t) = \nu(t) + K(t)[\beta(t) - z(t)].
\end{split}
\end{align*}
which is obtained by linearizing \eqref{eq:tracking_operator} about $\xi \in \TT$.
%
%
It can be shown that $D\PP(\xi)$ is itself a projection, so that $\zeta \in T_\xi\TT$
 if and only if $\zeta = D\PP(\xi) \cdot \zeta$.


\subsection{The PRONTO algorithm}
\label{subsec:pronto_alg}
Writing the cost in \eqref{eq:pronto_ocp} as the functional
\begin{align*}
  h(\xi) := \int_0^T \ell(x(\tau),u(\tau))\,\d\tau +m(x(T)),
\end{align*}
we see that the optimal control problem~\eqref{eq:pronto_ocp} is equivalent to 
the constrained optimization problem $\min_{\xi\in\TT} h(\xi)$.
Using the trajectory characterization 
and
defining $g(\xi) := h(\PP(\xi) )$ the constrained problem can be converted into
an unconstrained one as
%
$\min_{\xi \in \TT} h(\xi) = \min_{\xi} g(\xi).$

%

The PRONTO algorithm, stated in Algorithm~\ref{alg:PRONTO}, is based on a Newton
method applied to $\min_{\xi} g(\xi)$ and includes two key steps.  First, the
search direction $\zeta_i$ is determined by an optimization problem considering
the first and second derivatives of the nonlinear functional $g$.  Since the
derivatives of $g$ are computed, the projection $\PP$ is inherently considered
within the calculation of the search direction.  Moreover, the search direction
is limited to the tangent space of the trajectory manifold at the current trajectory
$\xi_i$, that is, $\zeta_{i}\in T_{\xi}\TT$.
Second, the update is performed using the projection $\PP$ in~\eqref{eq:pronto_alg_proj},
thus a feasible trajectory is determined after each iteration of the optimization algorithm.

\begin{algorithm}[H]
\begin{algorithmic}
\StatexIndent[0] \textsc{Given:} initial trajectory $\xi_0 \in \TT$

\StatexIndent[0] \textsc{For:} $i=0,1,2,\dots$
	\StatexIndent[0.5] redesign feedback $K$ if desired/needed 
	
	\StatexIndent[0.5] search direction
	\begin{equation}\label{eq:pronto_alg_desc_dir}
		\zeta_i=\argmin_{\zeta\in T_{\xi_i}\TT} \, Dg(\xi_i) \cdot \zeta+\textstyle\frac{1}{2}D^2g	(\xi_i)\cdot(\zeta,\zeta)
	\end{equation}
	\StatexIndent[0.5] step-size 
	\begin{equation*}
		 \gamma_i = \argmin_{\gamma\in (0,1]} \, g(\xi_i+\gamma\zeta_i) 
	\end{equation*}
	\StatexIndent[0.5] update
	\begin{equation}\label{eq:pronto_alg_proj}
		\xi_{i+1} = \PP(\xi_i+\gamma_i\zeta_i)
	\end{equation}
\end{algorithmic}

\caption{PRONTO}
\label{alg:PRONTO}

\end{algorithm}

\begin{remark}
Notice that step \eqref{eq:pronto_alg_desc_dir} consists of solving a
(standard) LQR problem in the form
\begin{align*}
  \begin{split}
  \underset{\zeta = (z(\cdot),v(\cdot))}{\minimize} \: & \: \int_0^T
  a(\tau)^T z(\tau) + b(\tau)^T v(\tau)
  \\ 
  &\: \: +
  \dfrac{1}{2}
  \begin{bmatrix}
    z(\tau) \\ v(\tau)
  \end{bmatrix}^T
  \begin{bmatrix}
    Q(\tau) & S(\tau)\\
    S(\tau)^T & R(\tau)
  \end{bmatrix}
  \begin{bmatrix}
    z(\tau) \\ v(\tau)
  \end{bmatrix} 
  \d\tau
  \\[0.2ex]
  &\: \: + z(T)^T P_1 z(T) + r_1^Tz(T)
  \\[1ex]
  \subject \: & \: 
  \dot{z} = A(t)z + B(t)v, \: \:\: z(0)=0.
  \end{split}
\end{align*}
Step~\eqref{eq:pronto_alg_proj} consists of computing the updated trajectory
$\xi_{i+1} = (x_{i+1}(\cdot), u_{i+1}(\cdot))$ by running the closed loop system
\eqref{eq:tracking_operator} with (given) curve
$(\alpha(\cdot), \mu(\cdot)) =\xi_i + \gamma_i \zeta_i = (x_i(\cdot) + \gamma_i
z_i(\cdot), u_i(\cdot) + \gamma_i v_i(\cdot))$. 
%
\oprocend
\end{remark}


\section{Final-state constrained PRONTO}
\label{sec:fs_pronto}
In this section, we introduce an optimization algorithm which solves the
nonlinear optimal state transfer problem.  The key approach is to: (i) introduce
a projection operator, inspired by the one introduced in~\cite{Hauser98} (and
recalled in Section~\ref{subsec:pronto}), such that not only the dynamics, but
also the terminal constraint is satisfied, and (ii) compute a descent direction
that satisfies the final-state constraint to first-order.

\subsection{Final-state constrained projection operator}
\label{subsec:fs_proj_oper}
The Projection Operator as recalled in Section~\ref{subsec:proj_oper} is not 
able to guarantee an exact matching of the terminal constraint. 
As a key step of our algorithm, we introduce a final-state constrained
projection operator,
$\xi = ( \alpha (\cdot), \mu (\cdot) ) \mapsto \PP_c (\xi)=\eta =
(x(\cdot),u(\cdot)),$
satisfying $x(T) = \alpha(T)$ where, as usual, $\xi$ is a curve while
$\eta \in \TT$ a trajectory.
%
%
%
Our idea is to \emph{design} the operator $\PP_c$ as an iterative routine in which, 
at each iteration: (i)~we perturb the actual trajectory in order to hit exactly the 
terminal constraint and (ii) we project the resulting curve by means of 
the standard projection operator~\eqref{eq:tracking_operator}.


The final-state constrained projection can be formalized in terms of an
infinite dimensional root-finding.
Given $x_T\in\real^n$, let us define a functional $\FF$ which associates to a
state-input curve $\xi=(\alpha(\cdot),\mu(\cdot))$ the difference between
its terminal state $\alpha(T)$ and $x_T$.  Hence, a \emph{trajectory} $\eta$
being a \emph{root} of $\FF$, i.e., such that $\FF(\eta) = 0$, is exactly what
we expect to be the result of the final-state constrained projection operator
$\PP_c$ when applied to a curve $\xi$.

Following the same high level idea in Section~\ref{subsec:pronto_alg} to derive
the PRONTO algorithm, we convert the constrained root-finding of $\FF$ into the
unconstrained root-finding of $\GG(\cdot) := \FF(\PP(\cdot))$, with $\PP$ being
the (unconstrained) projection operator introduced
in~\eqref{eq:tracking_operator}.


Given an initial curve $\xi$, the root of the functional $\GG$ is found by means
of an infinite-dimensional Netwon method.
Formally, at each iteration the perturbation $\zeta_k$ is obtained by setting to
zero the first order approximation of the perturbed functional, i.e. by solving
for $\zeta_k$ the following equation
\begin{equation}
\GG(\xi_{k}) + D\GG(\xi_{k}) \cdot \zeta_{k} = 0.
	\label{eq:root_finding_approx}
\end{equation}
%
%
%
Using the chain rule, the linear mapping $D\GG(\xi_{k})$ applied to a state-input 
curve $\zeta_{k}$ can be expressed as
%
  $D\GG(\xi_{k}) \cdot \zeta_{k} = D\FF(\xi_k)\cdot D\PP(\xi_k) \cdot \zeta_k$.
When $\xi_k$ is a trajectory, the linear mapping $D\PP(\xi_k)$ is a projection
on the tangent space $T_{\xi_k}\TT$ (see \cite{Hauser98}).  Moreover, the first
order expansion of the perturbed functional $\FF(\xi_k + \zeta)$ turns out to be
$D\FF(\xi_k)\cdot \zeta = (\pi_1\zeta)(T)$.
Thus, we can conclude that equation~\eqref{eq:root_finding_approx} simply
enforces a terminal condition on $\zeta_k$, i.e., find the state component
$z_k(\cdot)$ of $D\PP(\xi_k) \cdot \zeta_k \in T_{\xi_k}\TT$ such that
\begin{align}
x_k(T) - x_T + z_k(T) = 0.
  \label{eq:root_finding_terminal_constraint}
\end{align}
%
%
Note that, since the linear mapping $D\GG(\xi_{k})$ is not invertible, the
solution of~\eqref{eq:root_finding_approx} is not unique.

A finite dimensional counter-part of equation~\eqref{eq:root_finding_approx} is
a linear system of the form $M z + n = 0$. When $\ker M$ is non-empty, the
equation has not a unique solution. A typical approach to overcome this problem
is to consider the equivalent least-square problem, which selects the minimum
norm solution of the linear system.

Motivated by this finite-dimensional observation, a reasonable choice is to
select a $\zeta_{k}\in T_{\xi_{k}}\TT$ satisfying
condition~\eqref{eq:root_finding_terminal_constraint} with minimum $L_2$ norm.
It can be obtained solving the following linear quadratic optimal state transfer
problem
\begin{align*}
  \begin{split}
  \zeta_k := (z_k(\cdot),v_k(\cdot)) = \argmin_{(z(\cdot),v(\cdot))} & \,
  \frac{1}{2} \int_0^T\!
  \big \|z(\tau)\big \|^2 + \big \|v(\tau)\big \|^2 \, \d\tau
  \\
  \subj \: & \:
  \dot{z} = A(t) z + B(t) v
  \\
  & \: z(0)\!=\!0, \:\! z(T)\! =\! -x(T)\! +\! x_T,
  \end{split}
\end{align*}
where $A(\cdot)$ and $B(\cdot)$ result by the linearization of dynamics $f$ around the current iterate $\xi_k$.

A pseudo code of the constrained projection operator $\PP_c$ is given in the following
table (Algorithm~\ref{alg:cpo}). 

\begin{algorithm}[H]

\begin{algorithmic}
\StatexIndent[0] \textsc{Given}: a curve $\bar{\xi}$, a projection operator $\PP$ and a 

\StatexIndent[2.4] tolerance value {\tt\small tol}, \textsc{Set}: $\xi_0 = \bar{\xi}$

\StatexIndent[0] \textsc{For}: $k=0,1,2,\ldots$
  \StatexIndent[0.5] search direction
  \begin{align*}
    \zeta_k = 
    \argmin_{\zeta} &\: \textstyle \frac{1}{2}  \big\|\zeta \big\|_{L_2}^2 \\[1ex]
		\:\subj \: & \: \zeta \in T_{\xi_k}\TT \\
                          \: &\: (\pi_1\zeta) (T) = -\FF(\xi_{k})
  \end{align*}
  
\StatexIndent[0.5] update
	\begin{align*}
	\xi_{k+1} = \PP (\xi_k + \zeta_k)
	\end{align*}

\StatexIndent[0.5] \textsc{If}: $\| \FF(\xi_{k+1}) \| < {\tt tol}$, \textsc{Then:} break.\medskip

\StatexIndent[0] \textsc{Set:} $\PP_c(\bar{\xi}) =  \xi^*$, being $\xi^*$ the \emph{last} iteration trajectory.

\end{algorithmic}
\caption{Final-state constrained projection operator}
\label{alg:cpo}
\end{algorithm}

\begin{remark}
  The convergence of Algorithm~\ref{alg:cpo} can be guaranteed by satisfying the
  hypotheses of Newton-Kantorovich theorem (see, e.g., \cite{Kantorovich48,Ortega68}). \oprocend
\end{remark}

\subsection{fsPRONTO Algorithm}
\label{subsec:fspronto}

We are ready to present the final-state constrained PRojection Operator Newton
method for Trajectory Optimization (fsPRONTO) algorithm which is an iterative
algorithm able to solve problem~\eqref{eq:nonlin_state_transf}. 
%
The algorithm extends the PRONTO outlined in Section~\ref{subsec:pronto_alg} combining
a particular descent direction and the final-state constrained projection operator presented 
in Section~\ref{subsec:fs_proj_oper}.

First, we search for a descent direction $\zeta_i \in T_{\xi_i} \TT$ satisfying
the final constraint to first-order by means of a linear-quadratic state
transfer problem as in \eqref{eq:lin_state_transf}.
Since each $\xi_i$ is already feasible, in order to maintain feasibility to
first order, the perturbation $\zeta_i$ must satisfy the terminal constraint
$z_i(T) := (\pi_1 \zeta_i)(T) = 0$.
%
Second, we perform a backtracking line-search to modulate the descent direction.
Finally, we perform the projection step by means of the constrained projection operator described 
by Algorithm~\ref{alg:cpo}. 


The fsPRONTO algorithm is formally stated in the following table (Algorithm~\ref{alg:fsPRONTO}).

\begin{algorithm}[H]
\begin{algorithmic}

\StatexIndent[0] \textsc{Given:} initial trajectory $\xi_0$
	
\StatexIndent[0] \textsc{For:} $i=0,1,2,\dots$
  \StatexIndent[0.5] redesign feedback $K$ if desired/needed 

	\StatexIndent[0.5] constrained search direction
	\begin{align}
		\begin{split}
		\zeta_i = \argmin_{\zeta\in T_{\xi_i}\TT} & \: \, Dg(\xi_i)\cdot\zeta+\textstyle\frac{1}{2}D^2g	(\xi_i)\cdot(\zeta,\zeta)
		\\
		\subj \: & \: (\pi_1\zeta_i) (T) = 0
	\end{split}
	\label{eq:fsPRONTO_desc}
	\end{align}

	\StatexIndent[0.5] step-size
	\begin{equation}
		 \gamma_i = \argmin_{\gamma\in (0,1]} \, g(\xi_i+\gamma\zeta_i) 
	\label{eq:fsPRONTO_linesearch}
	\end{equation}

	\StatexIndent[0.5] constrained update
	\begin{equation}
		\xi_{i+1} = \PP_c(\xi_i+\gamma_i\zeta_i)
	\label{eq:fsPRONTO_proj}
	\end{equation}

\end{algorithmic}

\caption{Final-state constrained PRONTO.}
\label{alg:fsPRONTO}

\end{algorithm}

In the following, we have a closer look at some of the specific aspects of our 
newly presented Algorithm~\ref{alg:fsPRONTO}.

\begin{remark} 
  Notice that step \eqref{eq:fsPRONTO_desc} consists of solving a
  linear quadratic optimal state transfer problem in the form
\begin{align*}
  \begin{split}
  \underset{\zeta = (z(\cdot),v(\cdot))}{\minimize} \: & \: \int_0^T
  a(\tau)^T z(\tau) + b(\tau)^T v(\tau)
  \\ 
  & \: +
  \frac{1}{2}
  \begin{bmatrix}
    z(\tau) \\ v(\tau)
  \end{bmatrix}^T
  \begin{bmatrix}
    Q(\tau) & S(\tau)\\
    S(\tau)^T & R(\tau)
  \end{bmatrix}
  \begin{bmatrix}
    z(\tau) \\ v(\tau)
  \end{bmatrix} 
  \d\tau
  \\[1ex]
  \subject \: & \: 
  \dot{z} = A(t)z + B(t)v, \:\:\: z(0)=0, \:\: z(T) = 0,
  \end{split}
\end{align*}
as discussed in detail in Section~\ref{subsec:fscLQR}. 
Step~\eqref{eq:fsPRONTO_proj} consists of computing the updated trajectory
$\xi_{i+1} = (x_{i+1}(\cdot), u_{i+1}(\cdot))$ via Algorithm~\ref{alg:cpo} with a (given) curve 
$\bar{\xi} = \xi_i + \gamma_i \zeta_i = (x_i(\cdot) + \gamma_i z_i(\cdot), u_i(\cdot) + \gamma_i v_i(\cdot))$.
\oprocend
\end{remark}

\section{Numerical Computations}
\label{sec:simulations}
In this section we provide numerical computations showing the effectiveness of
the proposed nonlinear algorithm. We solve the optimal state transfer problem
for a driven inverted pendulum.
We consider the problem
\begin{align*}
  \underset{(x(\cdot),u(\cdot))}{\minimize} \: & \: 
  \int_0^T \textstyle\frac{1}{2} \big \|x(\tau)-x_d(\tau) \big \|^2_Q +\textstyle\frac{1}{2} \big \|u(\tau)-u_d(\tau) \big \|^2_R \,\d\tau
  \\
  \subject \:  & \: 
\left[\rule{0cm}{0.6cm}
  \begin{matrix}
    \dot{x}_1 \\[0.1cm] \dot{x}_2
  \end{matrix}
  \right]
  =
  \begin{bmatrix}
    x_2 \\
    \dfrac{g}{L}\sin x_1 - \dfrac{u}{L}\cos x_1
  \end{bmatrix}, \: 
	\begin{matrix} x(0) = x_0, \\[0.25em] x(T) = x_T,\end{matrix}
\end{align*}
%
with $L = 0.5$ m being the length of the pendulum and $g$ the gravity
acceleration.  We set the time horizon to $T=20$s.  Moreover,
$(x_d(\cdot), u_d(\cdot))$ is a (continuous) desired curve,
$Q\in\real^{2\times 2}$ is a symmetric, positive-definite matrix and $R$ is a
positive scalar.

%

Before testing the fsPRONTO algorithm, we highlight the applicability of the
final-state constrained projection operator presented in
Algorithm~\ref{alg:cpo}.  

\begin{figure*}[!ht]
  \centering
  \hspace{-0.2cm}
  \includegraphics[scale=0.4]{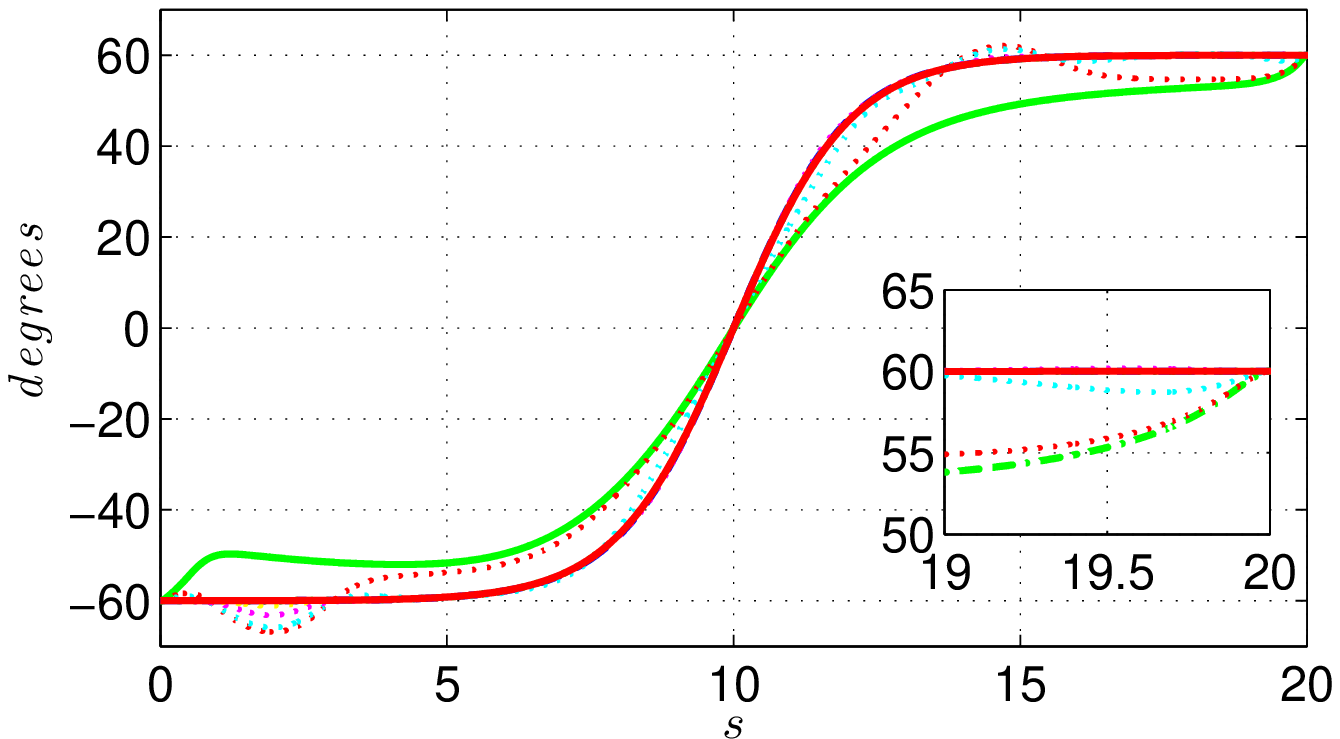}\hspace{-0.21cm}
  \includegraphics[scale=0.4]{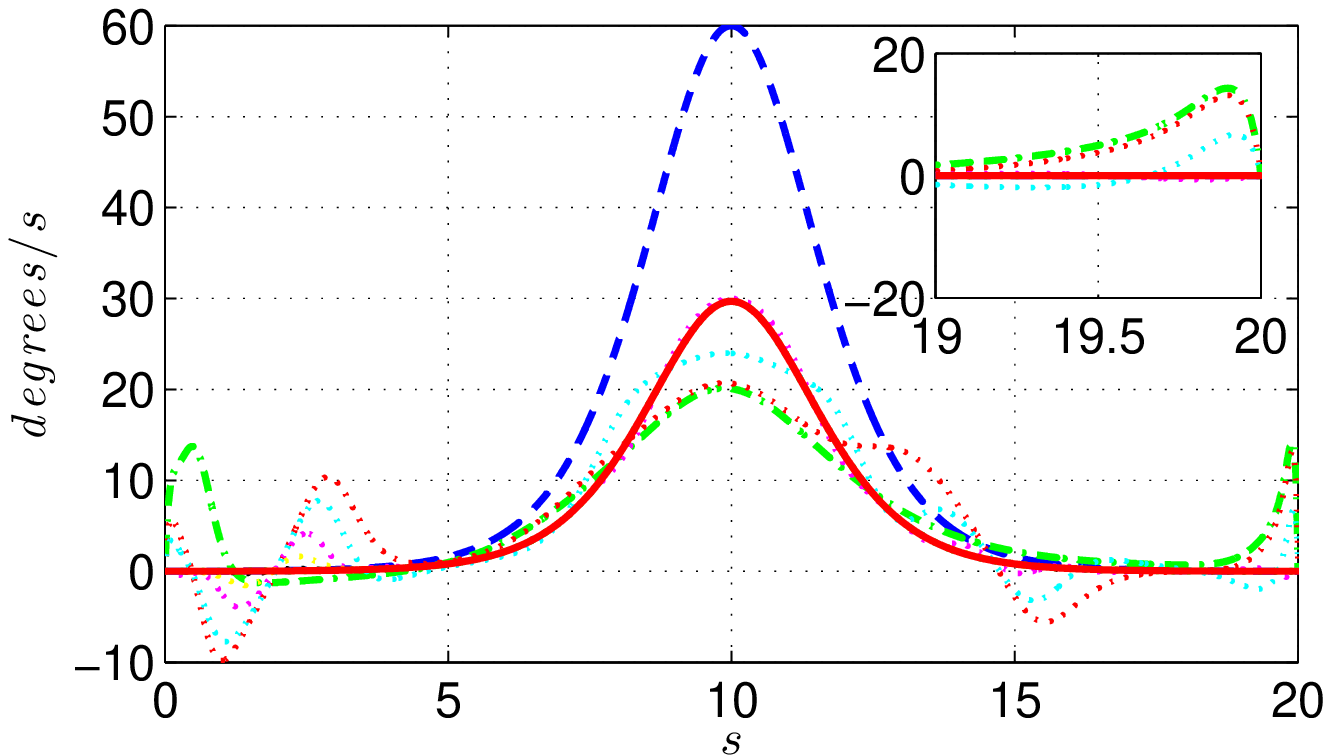}\hspace{-0.21cm}
  \includegraphics[scale=0.4]{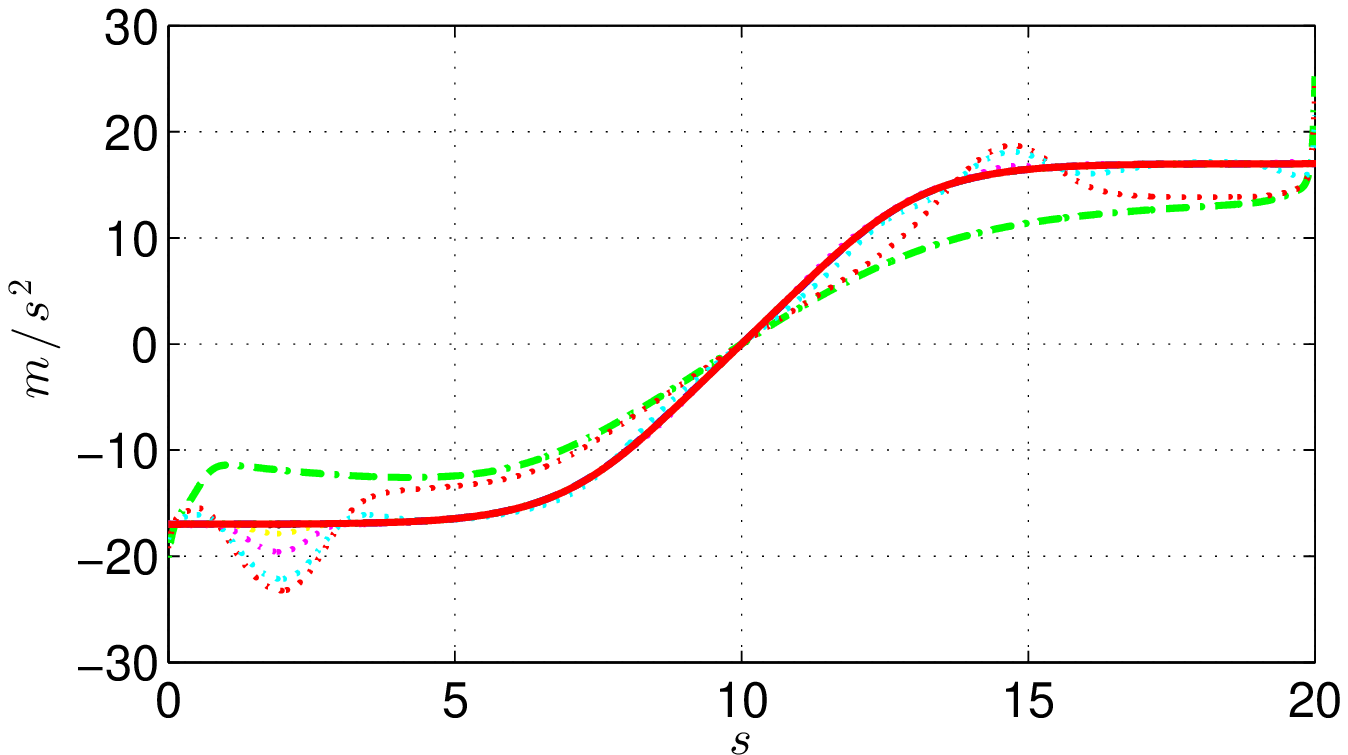}
  \caption{Evolution of $x_1$, $x_2$ and $u$ through the iterations of fsPRONTO
    (Algorithm~\ref{alg:fsPRONTO}). The desired curve (dashed blue), the initial
    (feasible) trajectory (dashed-dot green) and the optimal trajectory (solid
    red) are depicted. Intermediate (feasible) trajectories are plotted with
    light dotted lines.}
  \label{fig:fsPRONTO_iters}
	\vspace{-0.4cm}
\end{figure*}

\begin{figure}[!th]
  \centering
  \includegraphics[scale=0.4]{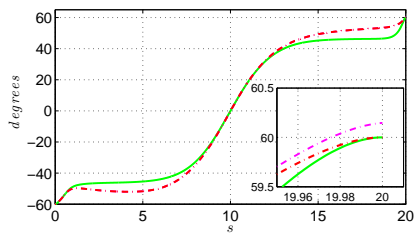}
  \caption{
    Final-state constrained projection operator: $x_1$ state
    component. Specifically, the unfeasible curve $\xi$ (solid green), the
    standard projection $\PP(\xi)$ (dashed-dot magenta) and the constrained
    projection $\PP_c(\xi)$ (dashed-dot red) are depicted.  }
  \label{fig:fs_proj_oper_iters_x1}
	\vspace{-0.4cm}
\end{figure}

We consider a given curve $\xi$ which is not a feasible trajectory of the inverted
pendulum.
The projected state $x_{1}$ is depicted in
Figure~\ref{fig:fs_proj_oper_iters_x1}. 
Both projections $\PP(\xi)$ (in magenta) and $\PP_{c}(\xi)$ (in red) provide a
trajectory close to the curve $\xi$ (in green). However, when closely checking
the terminal state, one can see that only the trajectory projected under
$\PP_c(\xi)$ satisfies the terminal constraint.

%

Next, we apply the fsPRONTO (Algorithm~\ref{alg:fsPRONTO}) in order to optimize
the trajectory of an inverted pendulum.  We use $Q = \diag(100,1)$ and $R=1$ as
cost parameters. The choice of a higher penalty on the first component $x_1$ of
the least-square distance will result in an optimal solution (solid red) which
almost overlaps the first component of the desired curve (dashed-dot blue) as
shown in Figure~\ref{fig:fsPRONTO_iters}.

It is worth nothing that, as expected, the algorithm guarantees recursive
feasibility. In fact, the terminal error, highlighted in the inset, is zero at
each iteration for both the state components.

%

\begin{figure}[!ht]
  \centering
  \includegraphics[scale=0.4]{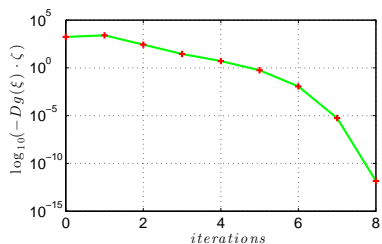}
  \caption{Convergence Rate of fsPRONTO Algorithm.}
  \label{fig:conv_rate}
\end{figure}

In Figure~\ref{fig:conv_rate} the descent at each iteration, in logarithmic
scale, is depicted. It gives a measure of the rate of convergence of the
algorithm which appears to be quadratic.

\section{Conclusions}
In this paper we have presented a new numerical approach for solving final-state
constrained optimal control problems.  The main advantage of the proposed method
is that it guarantees recursive feasibility of both the dynamics and the
final-state constraint at each iteration. Specifically, we have proposed a
Newton method, inspired to the one introduced in \cite{Hauser02}, based on: (i)
the design of a final-state constrained projection operator, being able to find
a trajectory satisfying the final constraint, and (ii) the computation of a
descent direction satisfying the final constraint to first-order.

\begin{small}
\bibliographystyle{IEEEtran}
\bibliography{cpronto}
\end{small}

\end{document}